\documentclass[conference]{IEEEtran}
\usepackage[utf8]{inputenc}
\usepackage[T1]{fontenc}
\usepackage{microtype}
\usepackage{graphicx}
\usepackage{dblfloatfix}
\usepackage{filecontents}
\usepackage[noadjust]{cite}
\usepackage{balance}
\usepackage{amsmath}
\usepackage{url}
\usepackage{csquotes}
\usepackage{mdframed}
\usepackage{subcaption}
\usepackage{multirow}
\usepackage{xcolor}
\usepackage{enumitem}
\usepackage[hidelinks]{hyperref}
\usepackage{tabularx}
\usepackage{tablefootnote}
\usepackage{marvosym}

\makeatletter
\newcommand{\textlabel}[2]{%
	\protected@edef\@currentlabel{#1}
	\phantomsection
	\label{#2}
}
\makeatother

\begin{document}
\title{Researcher or Crowd Member? Why not both!\\ The Open Research Knowledge Graph for Applying and Communicating CrowdRE Research}

\author{
\IEEEauthorblockN{Oliver Karras\IEEEauthorrefmark{1}, Eduard C. Groen\IEEEauthorrefmark{2}\IEEEauthorrefmark{3}, Javed Ali Khan\IEEEauthorrefmark{4}, Sören Auer\IEEEauthorrefmark{1}}

\IEEEauthorblockA{\IEEEauthorrefmark{1}Leibniz Information Centre for Science and Technology, Germany, \{oliver.karras, soeren.auer\}@tib.eu}

\IEEEauthorblockA{\IEEEauthorrefmark{2}Fraunhofer IESE, Germany, eduard.groen@iese.fraunhofer.de}

\IEEEauthorblockA{\IEEEauthorrefmark{3}Department of Information and Computing Sciences, Utrecht University, Netherlands}

\IEEEauthorblockA{\IEEEauthorrefmark{4}Department of Software Engineering, University of Science and Technology Bannu, Pakistan, engr\_javed501@yahoo.com}
}

\maketitle 

\begin{abstract}
In recent decades, there has been a major shift towards improved digital access to scholarly works. However, even now that these works are available in digital form, they remain document-based, making it difficult to communicate the knowledge they contain.
The next logical step is to extend these works with more flexible, fine-grained, semantic, and context-sensitive representations of scholarly knowledge. The Open Research Knowledge Graph (ORKG) is a platform that structures and interlinks scholarly knowledge, relying on crowdsourced contributions from researchers (as a crowd) to acquire, curate, publish, and process this knowledge.
In this experience report, we consider the ORKG in the context of Crowd-based Requirements Engineering (CrowdRE) from two perspectives: 
(1) As CrowdRE researchers, we investigate how the ORKG practically \textit{applies} CrowdRE techniques to involve scholars in its development to make it align better with their academic work. We determined that the ORKG readily provides social and financial incentives, feedback elicitation channels, and support for context and usage monitoring, but that there is improvement potential regarding automated user feedback analyses and a holistic CrowdRE approach.
(2) As crowd members, we explore how the ORKG can be used to \textit{communicate} scholarly knowledge about CrowdRE research. For this purpose, we curated qualitative and quantitative scholarly knowledge in the ORKG based on papers contained in two previously published systematic literature reviews (SLRs) on CrowdRE. This knowledge can be explored and compared interactively, and with more data than what the SLRs originally contained. Therefore, the ORKG improves access and communication of the scholarly knowledge about CrowdRE research.
For both perspectives, we found the ORKG to be a useful multi-tool for CrowdRE research.
\end{abstract}


\IEEEpeerreviewmaketitle

\section{Introduction}
Historically, research results were published in collections of printed works, such as journals~\cite{Auer.2020}. Studying the literature on a particular topic required hours or days of browsing through a library's collection. In recent decades, the research community has made great efforts to improve access to scholarly knowledge as part of the digital transformation \cite{Auer.2018, Jaradeh.2019}. Digitizing articles was the first step in this transformation. However, even digitized articles are just digital representatives of their printed counterparts~\cite{VanDeSompel.2009}. For this reason, digital articles impede scholarly communication by still being document-based~\cite{Bosman.2017}; while this representation is easy for humans to~process, it is poorly interlinked and not machine-actionable~\cite{Jaradeh.2019}. The next step in the digital transformation of scholarly knowledge requires a more flexible, fine-grained, semantic, and context-sensitive representation that allows humans to quickly compare research results, and machines to process them. This representation can hardly be created automatically, but requires domain experts~\cite{Auer.2020} and an infrastructure for acquiring, curating, publishing, and processing scholarly knowledge~\cite{Jaradeh.2019}.

Numerous projects \cite{Aryani.2017, Burton.2017, Bechhofer.2010, Hanson.2015, Sinha.2015, Hammond.2017, Meister.2017, Jaradeh.2019a} provide corresponding solutions using \textit{knowledge graphs} (see Section \ref{sec:background}) as structured, interlinked, and semantically rich representations of knowledge. While established knowledge graphs exist for representing encyclopedic and factual knowledge, e.g., in DBpedia \cite{Auer.2007} and WikiData~\cite{Vrandevcic.2014}, the use of knowledge graphs for scholarly knowledge is a rather new approach \cite{Auer.2018}. One of these projects is developing the \textit{Open Research Knowledge Graph}\footnote{\url{http://orkg.org/}} (ORKG)~\cite{Jaradeh.2019a}. The ORKG is a platform that uses crowdsourcing to acquire and curate scholarly knowledge. The project explores how scholarly knowledge can be acquired along the research lifecycle, relying on the manual acquisition and curation of scholarly knowledge through crowdsourced contributions from experts \cite{Jaradeh.2019, Jaradeh.2019a, Auer.2020}. Researchers from various research fields form a crowd that acquires, curates, publishes, and processes scholarly knowledge. Based on the publicly available beta version of the ORKG\footnotemark[1], the ORKG project team has two long-term goals. First, they aim to integrate more strategies for crowdsourcing to enable crowd members to contribute their research results to the ORKG in a more flexible and lightweight manner \cite[p. 9]{Jaradeh.2019a}. Second, the team aims to tailor the platform to the needs and requirements of the expert crowd by involving the crowd members in the development and soliciting feedback from them on problems and features \cite[p. 5]{Jaradeh.2019a}. Although these goals align with CrowdRE, the ORKG project team has not yet consciously applied CrowdRE. This fact aroused our interest to research the implementation of CrowdRE in the real development setting of the ORKG platform. In this experience report, we address two research questions:

\begin{mdframed}
	\textbf{RQ1:} What potential does the ORKG have as a platform for applying CrowdRE research in a real development setting?\vspace{0.2cm} \\
	\noindent
	\textbf{RQ2:} What potential does the ORKG have as a platform for communicating scholarly knowledge about CrowdRE research?
\end{mdframed}

Regarding RQ1, we take the perspective of \textit{CrowdRE researchers}. We describe the current state and features of the ORKG as a crowdsourcing platform that involves its crowd members in the development along the four key activities of CrowdRE (\textit{motivating crowd members}, \textit{eliciting feedback}, \textit{analyzing feedback}, \textit{monitoring context \& usage data})~\cite{Groen.2017}. This overview shows to what extent and how the ORKG already addresses CrowdRE, and highlights improvement potential. CrowdRE researchers benefit from this overview as it enables them to assess the suitability of the ORKG as a basis for their future work and studies on CrowdRE research, while the efficacy of CrowdRE in a real development setting can be demonstrated to practitioners. Regarding RQ2, we take the perspective of \textit{crowd members}. We provide insights into our experiences with acquiring, curating, and publishing qualitative and quantitative scholarly knowledge about CrowdRE research in the ORKG. Using two examples from CrowdRE research~\cite{Santos.2019a, Khan.2019}, we illustrate how (CrowdRE) researchers can use the ORKG to improve access and communication of scholarly knowledge about their research. Based on the two perspectives, we assess the potential of the ORKG as a platform for applying and communicating CrowdRE research. We gained the following insights:

\begin{mdframed}
The ORKG has a two-fold potential for CrowdRE:
	\begin{enumerate}[leftmargin=0.18cm]
	   	\item[] (1) The platform and its features provide a solid basis for applying CrowdRE research in a real development setting and in close collaboration with the ORKG project~team.
		
		\item[] (2) The platform enables the communication of CrowdRE research by allowing more comprehensive curation of scholarly knowledge than document-based works do.
	\end{enumerate}
	This two-fold potential makes the ORKG a useful multi-tool for applying and communicating CrowdRE research.
\end{mdframed}


\section{Background}
\label{sec:background}
In this section, we briefly introduce \textit{knowledge graphs} and their contribution to the digital transformation of scholarly communication towards graph-based knowledge sharing.
According to Brack et al. \cite[p. 1]{Brack.2020}, \enquote{\textit{A knowledge graph} (\textit{KG}) \textit{consists of} (1) \textit{an ontology describing a conceptual model} (\textit{e.g., with classes, relation types, and axioms})\textit{, and} (2) \textit{the corresponding instance data }(\textit{e.g., objects, literals, and} $\langle$\textit{subject, predicate, object}$\rangle$\textit{-triplets}) \textit{following the constraints posed by the ontology} (\textit{e.g., instance-of relations, axioms, etc.})\textit{. The construction of a KG involves ontology design and population with instances.}}

In the context of scholarly communication, a knowledge graph represents original research results semantically, i.e., explicitly and formally, and comprehensively links existing data, metadata, knowledge, and information resources \cite{Auer.2018a}.

The organization of scholarly communication and knowledge based on the structured, standardized, and semantic representation form of a knowledge graph offers several benefits~\cite{Auer.2018a}. 
It \textit{increases the unique identification} of the relevant artifacts, concepts, attributes, and relationships of research results. All these elements can be linked with each other, considering the constraints imposed by the underlying ontology. 
It \textit{increases traceability} through improved and explicit linking of the artifacts and information sources. In this way, graph-based knowledge sharing also helps to \textit{reduce redundancy and duplication} because repetitive content, such as related work on a topic, can be continuously described, stored, and communicated over time. This type of scholarly communication curbs the proliferation of scientific publications and leads to an \textit{increase in efficiency} by avoiding media discontinuities in the various phases of scientific work. 
It \textit{reduces ambiguity} through more terminological and conceptual clarity because the concepts and relationships can be reused across disciplinary boundaries. 
It \textit{increases machine actionability} on the content of  scientific publications, enabling machines to understand the structure and semantics of the content of scientific publications. 
Finally, because the content is machine-actionable, it \textit{increases development opportunities} for applications that provide search, retrieval, mining, and assistance features for scholarly knowledge. These applications could support open science by making knowledge more accessible, e.g., to early career scientists, lay people, or researchers with a visual impairment.

\section{Related Work}
\label{sec:related-work}
In recent years, the role of a crowd has attracted much attention across many disciplines through means such as crowdsourcing. Numerous projects have emerged offering platforms that involve a crowd as an inherent part of their systems \cite{Huws.2016, Mao.2017, Hosseini.2014}.
The digital transformation has compelled the domain of requirements engineering (RE) to respond in various ways, including placing greater emphasis on automated RE \cite{Villela.2018}. A particular kind of data-driven RE \cite{Maalej.2015a, Maalej.2019} relies on the involvement of a crowd as a source of data, and is typically referred to as CrowdRE~\cite{Groen.2017}. CrowdRE describes an iterative cycle of eliciting feedback from the crowd and monitoring context and usage data to derive the needs and requirements of the crowd, which, once validated, are implemented into the product~\cite{Groen.2017}. Current CrowdRE research focuses on the use of various existing channels with a general purpose, aimed at accessing the crowd, such as social networks and mobile application marketplaces~\cite{Khan.2019}. These independent channels are separated from the actual product but allow crowd members to provide feedback on the product. However, they are not enough for successful CrowdRE \cite{Stade.2017, Khan.2019}; a combined feedback and monitoring solution needs to be integrated into the actual product to optimally support CrowdRE. This integration can simplify crowd involvement by making it easier for crowd members to provide feedback and thus actively participate in the development. It also helps developers better understand the feedback given because of the details obtained through the monitored context and usage data \cite{Stade.2017, Khan.2019}. Approaches such as FAME \cite{Oriol.2018} propose a possible integration of multiple data sources. The crowd can also be involved through crowdsourced tasks, e.g., requirements classification \cite{VanVliet.2020}. In the long term, however, we need systematic and holistic approaches covering the entire software development process to create information systems that involve a crowd operationally, e.g., to provide content or form a social network, and to support development with user feedback. As this topic has only emerged in recent years and is still being researched, such approaches are still lacking. To address this deficit, platforms and projects must work closely with researchers to implement CrowdRE in real development settings \cite{Stade.2017, Khan.2019, Franch.2020}.

\section{The ORKG as a Platform for CrowdRE}
\label{sec:orkg_as_a_platform}
In his CrowdRE'19 keynote, Glinz~\cite{Glinz.2019} emphasized the need for CrowdRE to venture out into open source and open research settings because its characteristics make it highly suitable for their respective crowds. Based on this notion, the ORKG aroused our interest for several reasons. (1) The ORKG is an open source platform for open research. (2) The goals of the ORKG project team are aligned with CrowdRE. (3) The project team seeks to collaborate with others on use cases and new features for the ORKG\footnote{\url{https://projects.tib.eu/orkg/get-involved/}}. (4) The project offers two perspectives on the ORKG: As researchers for applying CrowdRE research in a real development setting, and as crowd members for communicating this research. We first describe our analysis procedure in Section \ref{sec:platform_procedure}. We then present our assessment of the features of the ORKG as a crowdsourcing platform that involves its crowd members in the development in Section \ref{sec:platform_features}, and we report our experiences in using the ORKG in Section \ref{sec:platform_experience}.

\subsection{Procedure}
\label{sec:platform_procedure}
\begin{figure}[htbp]
    \captionsetup{justification=justified}
	\centering
	\includegraphics[width=0.9\columnwidth]{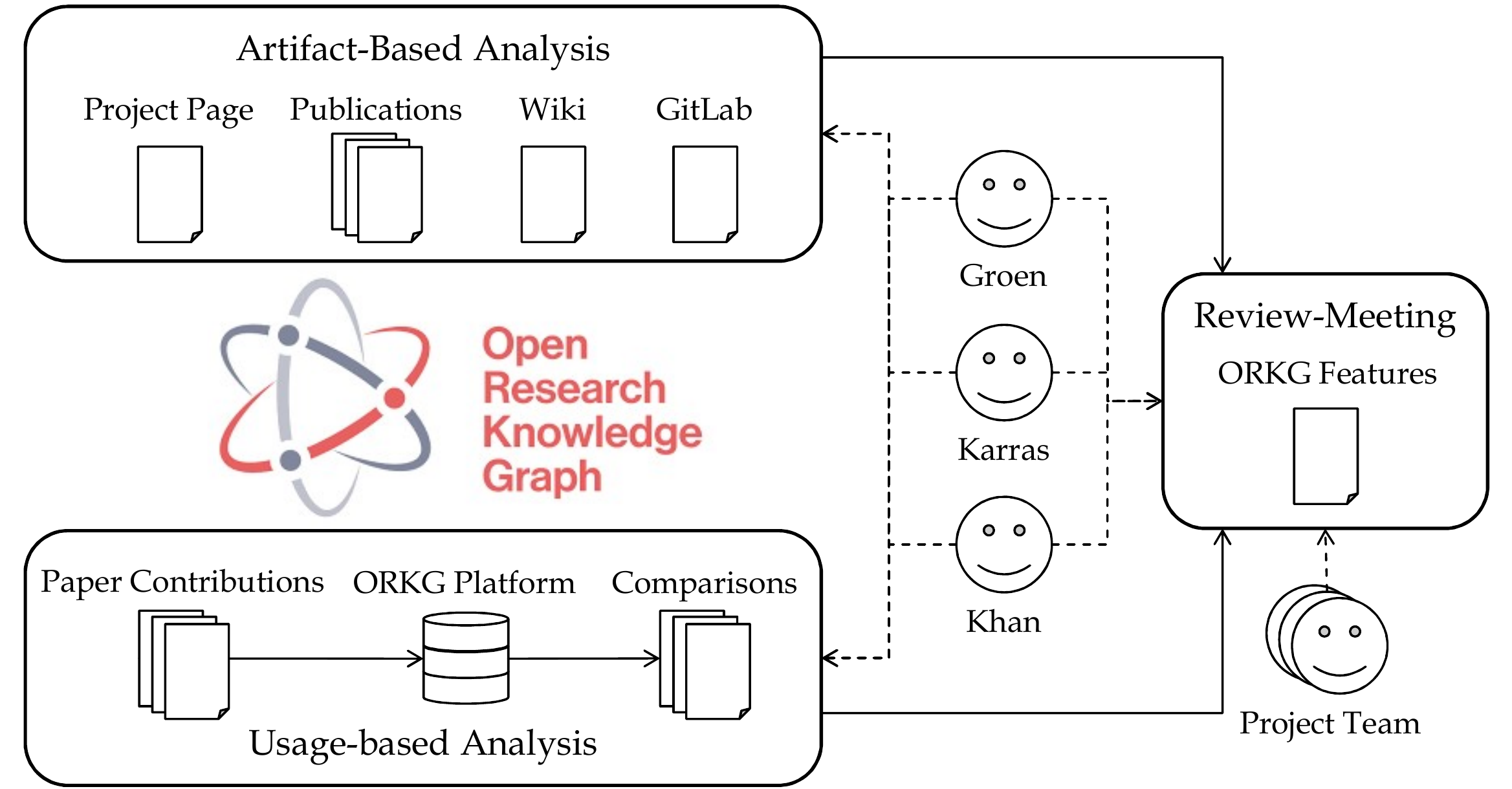}
	\caption{Overview of the procedure for analyzing the ORKG.}
	\label{fig:fig1}
	\vspace{-0.4cm}
\end{figure}

We performed an artifact- and usage-based analysis of the ORKG (see \figurename{ \ref{fig:fig1}}). On the one hand, we analyzed the documentation from the project page\footnote{\url{https://projects.tib.eu/orkg/}}, the wiki\footnote{\url{https://gitlab.com/TIBHannover/orkg/orkg-frontend/-/wikis/home}}, GitLab\footnote{\url{https://gitlab.com/groups/TIBHannover/orkg/-/issues}}, and related publications \cite{Auer.2018, Jaradeh.2019, Brack.2020, Auer.2020, Jaradeh.2019b, Oelen.2020, Wiens.2020}. On the other hand, we used the ORKG as crowd members to acquire, curate, and publish scholarly knowledge about CrowdRE research~\cite{Karras.2021, Karras.2021b}. From the analysis, we identified features that typify the ORKG as a crowdsourcing platform. We structured these features according to Hosseini et al.'s \cite{Hosseini.2014} reference model for crowdsourcing\textemdash the so-called \textit{four pillars of crowdsourcing}. This reference model provides a taxonomy of crowdsourcing to describe the individual features of the four pillars\textemdash the \textit{crowd}, the \textit{crowdsourcer}, the \textit{crowdsourced task}, and the \textit{crowdsourcing platform}\textemdash in a hierarchical manner. When we presented our results to the ORKG project team during a review meeting for verification, they provided suggestions to extend some descriptions. \tablename{ \ref{tbl:t1}} shows an excerpt\footnote{The full overview is available online in a supplement to this paper \cite{Karras.2021a}.} of the features of the ORKG as a crowdsourcing platform. 

\begin{table*}[!htb]
    \captionsetup{justification=justified}
	\centering
	\caption{Excerpt of identified features of the ORKG as a crowdsourcing platform (cf. Karras et al. \cite{Karras.2021a})
	}
	\label{tbl:t1}
    \begin{tabularx}{\textwidth}{|l|X|}
    \hline
    \textbf{Feature} & \textbf{Description} \\ \hline
    \multicolumn{2}{|l|}{\textbf{Pillar 1: The crowd}} \\ \hline
    \multicolumn{2}{|l|}{5.4 Motivation} \\ \hline
    5.4.1 Mental satisfaction & The crowd members support open data, open research, and open knowledge for all. \\ \hline
    5.4.2 Self-esteem & The crowd members know that they support the research community. \\ \hline
    5.4.3 Personal skill development & The crowd members can develop their research skills by creating state-of-the-art comparisons and smart reviews. \\ \hline
    5.4.4 Knowledge sharing & The crowd members share their research by acquiring and curating their scholarly knowledge with others. \\ \hline
    5.4.5 Love of community & The crowd members value each other's results since the platform addresses an open research community. \\ \hline
    
    \multicolumn{2}{|l|}{\textbf{Pillar 2: The crowdsourcer}} \\ \hline
    \multicolumn{2}{|l|}{1. Incentives provision} \\ \hline
    1.1 Financial incentives & The project team launched the ORKG Curation Grant Competition in May 2021, paying €400 per month for regular contributions to the ORKG (initially limited to six months). \\ \hline
    1.2 Social incentives & The crowdsourcer uses public acknowledgments of contributors and curators of scholarly knowledge on the platform with prominently visible rankings and mentions. \\ \hline
    1.3 Entertainment incentives & This feature is not currently supported. \\ \hline
    
    \multicolumn{2}{|l|}{\textbf{Pillar 3: The crowdsourced task}} \\ \hline
    7.1 Problem solving & The task of acquiring, curating, publishing, and processing scholarly knowledge can consider a specific research problem that can be answered with an analysis of a state-of-the-art comparison in the respective research field. \\ \hline
    7.2 Innovation & The task of acquiring, curating, publishing, and processing scholarly knowledge can lead to new ideas. \\ \hline
    7.3 Co-creation & The task of acquiring, curating, publishing, and processing scholarly knowledge requires collaboration with crowd members to communicate and maintain scholarly knowledge in the long term. \\ \hline
    
    \multicolumn{2}{|l|}{\textbf{Pillar 4: The crowdsourcing platform}} \\ \hline
    \multicolumn{2}{|l|}{1. Crowd-related interactions} \\ \hline
        
    1.9 Provide feedback loops & 
    The platform uses several options to provide feedback to the crowd as a whole: mailing list\footnotemark[2], Twitter account\footnotemark[7], project page\footnotemark[3], and the ORKG website\footnotemark[1].
    They are used to communicate regularly about the current status and changes to the platform, including technical improvements and content development achieved through crowdsourcing. However, there are currently no mechanisms to provide feedback to individual crowd members.\\ \hline
    \multicolumn{2}{|l|}{2. Crowdsourcer-related interactions} \\ \hline
    2.8 Provide feedback loops & The platform gives the crowd several options to provide feedback to the project team as the crowdsourcer:
    \begin{enumerate}
        \item Different communication mechanisms for contacting the project team:
        
        Chatwoot\footnotemark[8], email contacts\footnotemark[2], Skype group\footnotemark[2], Twitter account\footnotemark[7], and a GitLab issue tracker\footnotemark[5].
        
        \item Surveys integrated into the platform after specific processes, e.g., adding a publication to the platform.
        \vspace{-0.32cm}
    \end{enumerate} \\ \hline
    \multicolumn{2}{|l|}{3. Task-related facilities} \\ \hline
    3.3 Store history of completed tasks & The platform stores all tasks and changes for each individual crowd member. In addition, the platform supports versioning for created comparisons and smart reviews. \\ \hline
    \multicolumn{2}{|l|}{4. Platform-related facilities} \\ \hline
    4.3 Provide ease of use & The platform has its own front-end development team for continuously improving the interface for the crowd. \\ \hline
    4.4 Provide attraction & The platform has its own front-end development team for continuously improving the interface for the crowd. \\ \hline
    \end{tabularx}
    \vspace{-0.2cm}
\end{table*}

\subsection{Identified Features of ORKG as a Crowdsourcing Platform}
\label{sec:platform_features}
Described along the four pillars of crowdsourcing \cite{Hosseini.2014}, the ORKG is a \textit{crowdsourcing platform} with a \textit{crowd}, mainly consisting of researchers, that is diverse in terms of spatial distribution, gender, age, and expertise. As of August 2021, the crowd consists of 530 members, 307 of whom actively contribute scholarly knowledge to the platform. The ORKG project team as the sole \textit{crowdsourcer} focuses on expert-based crowdsourcing. Nevertheless, the platform is open to anyone willing to give it a try. The \textit{crowdsourced task} is the acquisition and curation of scholarly knowledge by publishing and processing state-of-the-art comparisons and corresponding articles\textemdash so-called \textit{smart reviews}\textemdash in arbitrary research fields. To better reflect the platform's use of CrowdRE from a researcher's perspective, we organized the identified features of the platform along the four key activities of CrowdRE: \textit{motivating crowd members}, \textit{eliciting feedback}, \textit{analyzing feedback}, and \textit{monitoring context \& usage data} \cite{Groen.2017}.\vspace{0.2cm}

\subsubsection{Motivating Crowd Members}
Crowd members must be motivated to become and remain active participants. Their motivation can be intrinsic and/or extrinsic. \textit{Intrinsic motivation} is mainly rooted in \textit{knowledge sharing} and \textit{love of the community}. This motivates crowd members to share their research by acquiring and curating scholarly knowledge on the platform with others, while valuing the contributions of the other crowd members. In this way, crowd members also achieve \textit{mental satisfaction}, boosted \textit{self-esteem}, and \textit{personal skill development} since they support the research community with open data, open research, and open knowledge by creating state-of-the-art comparisons and writing smart reviews. \textit{Extrinsic motivation} is mainly achieved through \textit{social incentives} in the ORKG. The project team essentially uses public acknowledgments of crowd members to motivate them, through visible rankings and citable contributions, to contribute scholarly knowledge to the ORKG. Among other things, a unique Digital Object Identifier (DOI) can be assigned to a comparison or a smart review. 
In May 2021, the project team also introduced \textit{financial incentives} through the ORKG Curation Grant. Crowd members can apply for this grant based upon demonstrable contributions to the ORKG. If accepted, they receive €400 per month and commit to making regular contributions to the ORKG for a period of six months. At the moment, the ORKG does not provide \textit{entertainment incentives}, such as gamification.

\begin{mdframed}
	\textlabel{\protect Finding~1}{f1}\textbf{Finding 1:}
	The ORKG strongly relies on the intrinsic motivation of researchers to share and communicate research contributions. The platform does support extrinsic motivation by providing social incentives through public acknowledgments of contributions, which are also citable through DOIs. Recently, the project team added a financial incentive. The ORKG currently does not employ entertainment incentives such as enjoyment, fun, and gamification.
\end{mdframed}

\footnotetext[7]{\url{https://twitter.com/orkg_org}}
\footnotetext[8]{\url{https://www.chatwoot.com/}}

\subsubsection{Eliciting Feedback}
The foundation of CrowdRE is the ability to elicit feedback from the crowd and derive requirements in return. The ORKG already provides several \textit{feedback loops} through which the crowd can provide feedback to the project team, including means of contacting the project team. (1) The ORKG uses Chatwoot; a support communication system that can be integrated into platforms such as websites to enable direct communication between crowd members and administrators. (2) The project team provides email contacts and a Skype group to offer help and support. Both channels are less typical for CrowdRE, but can certainly be used as data sources for analysis. (3) The ORKG has a Twitter account and a GitLab issue tracker, both of which are common data sources for eliciting and analyzing user feedback. (4) The project team also started to integrate feedback loops into the ORKG through surveys conducted after certain activities in order to understand how well crowd members got along with the system.

Besides feedback loops from the crowd to the project team, the ORKG also offers mechanisms in the opposite direction. These mechanisms include a mailing list, the Twitter account, the project page, and the ORKG platform itself. These mechanisms are used to communicate regularly about the current status of and changes to the ORKG, including its technical and content development that was achieved with the help of the crowd. However, the ORKG lacks mechanisms for providing targeted feedback to individual crowd members.

\vspace{2mm}

\begin{mdframed}
	\textlabel{\protect Finding~2}{f2}\textbf{Finding 2:}
	The ORKG project team uses various mechanisms to communicate bilaterally with the crowd. Channels external to the ORKG include Twitter and a GitLab issue tracker, while integrated channels include Chatwoot, surveys, Skype, and email. In this way, the ORKG uses both familiar and less typical feedback channels for CrowdRE, and is in the process of integrating feedback mechanisms into the platform itself.
\end{mdframed}

\subsubsection{Analyzing Feedback}
A central concern in CrowdRE is the derivation of requirements from the elicited feedback through analysis. The ORKG project team currently relies on direct communication with the crowd, and thus on \textit{immediate analysis and processing of the feedback}, which is reflected in the extensive use of direct communication channels such as Chatwoot and a Skype group. Accordingly, the analysis of feedback is done manually, without any (semi-)automated analyses. This was a conscious design decision due to the smaller size of the project team and the expected limited\textemdash thus, manageable\textemdash crowd size in the early stages of the project. In the long term, however, more (semi-)automated CrowdRE feedback analysis measures are to be put in place if the platform is to cater for the anticipated thousands of researchers from various research fields, where direct contact between crowd members and the project team will be limited by necessity.

\begin{mdframed}
	\textlabel{\protect Finding~3}{f3}\textbf{Finding 3:}
	The ORKG project team analyzes the elicited feedback manually and immediately. It deliberately has no (semi-)automated feedback analysis approaches in place. Once the crowd's size is no longer manageable, keeping the crowd involved in the development inevitably demands other feedback analysis paradigms.
\end{mdframed}

\subsubsection{Monitoring Context \& Usage Data}
The ORKG project team provides several options to assist the crowd members in using the ORKG. These options include tutorial videos, a guided tour of and tooltips on the user interface, templates, comprehensive documentation, and different communication mechanisms for support. The project's dedicated front-end development team continuously improves the user interface for the crowd based on the provided feedback, which can be enriched with monitored context and usage data to better understand the feedback of the crowd members. For usage monitoring, the ORKG stores the history of all changes and tasks completed by the crowd members, and the web analytics tool Matomo\footnote[9]{\url{https://matomo.org/}}
is used to evaluate the crowd members' journeys on the platform. In addition, administrators and curators can supervise the activities of the crowd in the background.
\vspace{-0.1cm}
\begin{mdframed}
	\textlabel{\protect Finding~4}{f4}\textbf{Finding 4:}
	 The ORKG has basic approaches for monitoring context and usage data in place. It was inherently designed to store any history, which fundamentally facilitates tracking usage behavior in compliance with data privacy regulations. This historical data can be supplemented with web analytics data collected through Matomo.
\end{mdframed}

\subsection{Experiences with Using ORKG as Crowd Members}
\label{sec:platform_experience}
Besides reflecting on the ORKG as CrowdRE researchers, we also experienced the platform and its features as crowd members. We created two state-of-the-art comparisons in the ORKG \cite{Karras.2021, Karras.2021b} based on papers contained in two previously published systematic literature reviews (SLRs) on CrowdRE~\cite{Khan.2019, Santos.2019a}. We selected these for three reasons. (1)~One author of each SLR co-authored this paper, enabling us to ensure that the contributions from the SLRs' papers were curated according to the interpretation of its authors. (2)~Both SLRs represent milestones regarding the current status of two important topics in CrowdRE research. (3) Due to their different nature, the SLRs allow us to determine the potential of the ORKG for qualitative and quantitative data, respectively.\vspace{0.2cm}

\subsubsection{Case~I -- Crowd Intelligence in Requirements Engineering}
The SLR conducted by Khan et al. \cite{Khan.2019} marks one of the most comprehensive overviews of the literature on CrowdRE to date, encompassing 77 papers. A noteworthy contribution of this SLR is that it organizes these papers according to five phases of RE, along with the CrowdRE utilities applied in each paper, such as the crowd, the crowdsourced task, incentives for motivation, and channels for feedback elicitation. This contribution is based on a qualitative expert analysis of the papers. Although the classification provides a strong overview of the works published until 2019, it is nearly impossible to dynamically keep this overview up-to-date in the long term as a document-based publication.\vspace{0.2cm}

\subsubsection{Case~II -- User Feedback Classification Approaches}
The SLR conducted by Santos et al. \cite{Santos.2019a} provides a quantitative comparison of 43 papers on how well Machine Learning (ML) algorithms perform in classifying elicited user feedback. Out of a total of 78 classification categories in the field of feedback analysis for RE \cite{Santos.2019}, their SLR made a quantitative comparison for only the single most frequently found classification category, ``Feature Request'', because the large tables needed to present all results conflicted with space constraints. Realistically, a publication that presents such a comparison even for a subset of the 78 classification categories could potentially become long and repetitive, or require complex groupings of classification categories. The presentation would then likely overshoot its goal of conveying comparable scholarly knowledge. \vspace{0.2cm}

\begin{figure*}[!t]
    \captionsetup{justification=justified}
	\centering
	\includegraphics[width=0.9\textwidth]{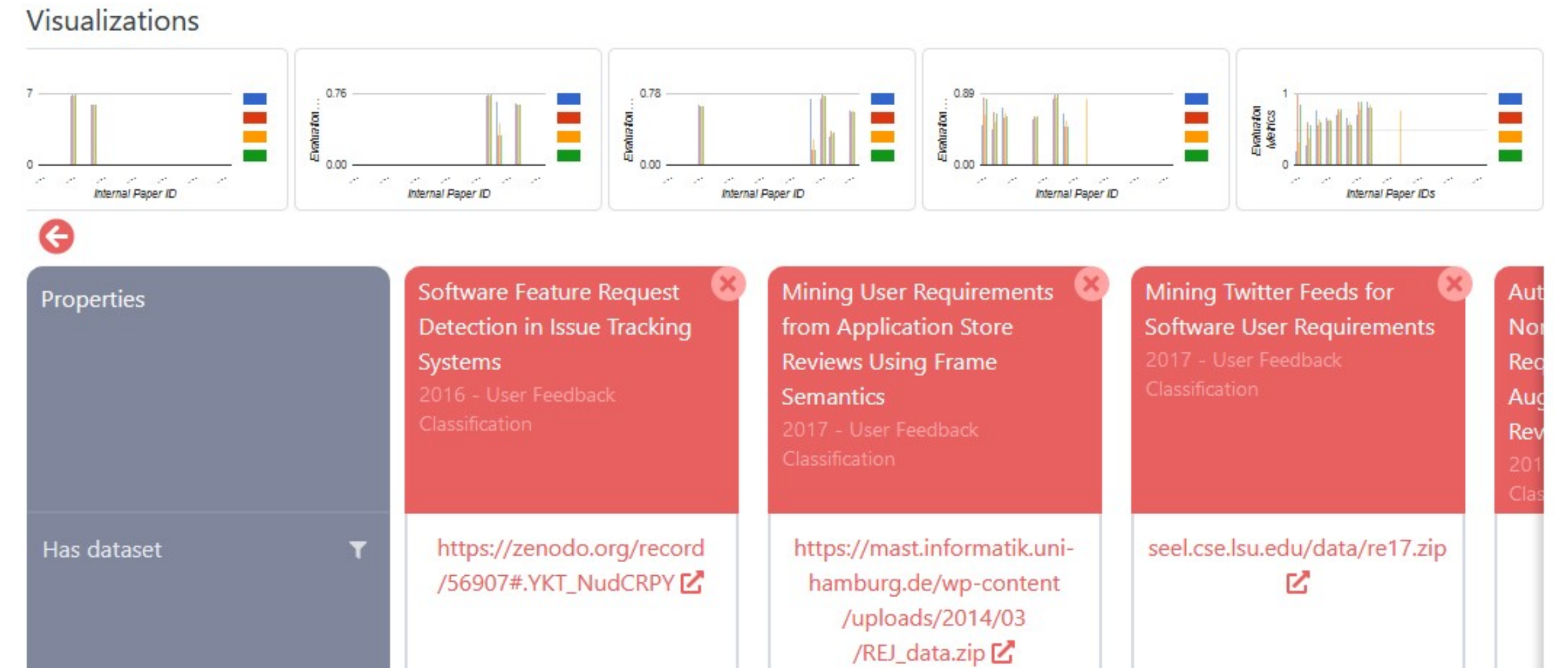}
	\caption{Excerpt from our comparison for Case~II \cite{Karras.2021}.}
	\label{fig:fig3}
	\vspace{-0.5cm}
\end{figure*}

\subsubsection{State-of-the-Art Comparisons with the ORKG}
Cases~I and II provide valid motivations for creating state-of-the-art comparisons with the ORKG, given the need for continuous curations of scholarly knowledge as well as flexible, fine-grained, semantic, and context-sensitive representations.

The ORKG organizes the acquired and curated scholarly knowledge by paper as a collection of so-called \textit{contributions}, which address a research problem and consist of scholarly knowledge. This knowledge is stored in a knowledge graph, from which the crowd members distill the contributions they are looking for. The selected contributions are compared in state-of-the-art comparisons. \figurename{~\ref{fig:fig3}} shows an excerpt from our comparison for Case~II~\cite{Karras.2021}. The columns denote the contributions by paper, and the rows denote the scholarly knowledge. The comparison of Case~II currently describes the ML classifiers, ML features, and the quantitative classifier performance values (Precision, Recall, the $F_1$ and $F_\beta$ measures, and Berry's \cite{Berry.2021} task-based $\beta_T$ value) from the 19 papers described in the SLR by Santos et al.~\cite{Santos.2019a} that used the classification category ``Feature Request''. The comparison of Case~I so far consists of contributions from 27 of the 77 papers from the SLR by Khan et al.~\cite{Khan.2019}, describing the relation of the papers to five phases of RE and the CrowdRE utilities applied~\cite{Karras.2021b}. We are still in the process of adding the contributions from the remaining 50 papers, which is more time-consuming than for the quantitative data from Case~I because of the expert judgments needed for classifying the papers' contributions. The comparison of the 27 papers makes it easy to identify, for example, the four papers that address the runtime purpose of monitoring for requirements evolution.

With the created comparisons \cite{Karras.2021,Karras.2021b}, we achieved our goal of acquiring and curating the detailed results of both SLRs. The knowledge-based representation in the form of comparisons has several advantages over a purely document-based representation. The comparisons are interactive and allow filtering of views by different scholarly knowledge contained in each row, even by specific value ranges of qualitative and quantitative content. The ORKG also provides a service for generating several graphical visualizations based on data in the comparisons, helping the reader understand information faster than through the large comparison table. The most important feature of the ORKG is that the added contributions and created comparisons are available to anyone. In this way, every crowd member can use the curated scholarly knowledge and created comparisons as a basis for new comparisons. Moreover, the existing comparisons can be expanded with additional scholarly knowledge from papers already included, and even with new contributions from papers added later to the ORKG. We already expanded several contributions, e.g., the results of other classifications reported in Dhinakaran et al.'s paper~\cite{Dhinakaran.2018}\footnote[10]{\url{https://www.orkg.org/orkg/paper/R76818/R76825}}. For Case~I, we added the details of the three crowd properties \textit{scale}; \textit{level of knowledge, skills \& expertise}; and \textit{roles}, which are only briefly and superficially described in the SLR \cite{Khan.2019}. For Case~II, we added links to the datasets used and performance values to classification categories other than ``Feature Request''. This expansion is relevant to enable long-term curation. For example, a development succeeding the SLR by Santos et al.~\cite{Santos.2019a} are reports of Deep Learning algorithms showing promising results in classifying user feedback~\cite{Reddy.2021,Stanik.2021}, which should be successively added to the comparison.

Despite all these advantages, the ORKG also has limitations. Most of the limitations we experienced can be attributed to the development status of the platform, which is currently in beta. Further development of the ORKG must improve interactions for the expert crowd by enabling better workflows for entering data and creating visualizations. Nevertheless, we also experienced that the project team has always responded directly to our reported issues, which we could see getting added to the GitLab issue tracker\footnote[11]{\url{https://gitlab.com/TIBHannover/orkg/orkg-frontend/-/issues/634}} and addressed shortly thereafter.

\section{Discussion}
\label{sec:platform_discussion}
The ORKG aroused our interest as a crowdsourcing platform for applying and communicating CrowdRE research. In this experience report, we explored whether the ORKG can promote the potential of CrowdRE in open source and open research settings, taking two perspectives: that of CrowdRE researchers and that of crowd members.

Our first contribution is that we provide a comprehensive overview of the ORKG's features as a crowdsourcing platform for acquiring and curating scholarly knowledge \cite{Karras.2021a}, mapped to the four key activities of CrowdRE. Our findings show that the ORKG is a crowdsourcing platform offering several features that can facilitate successful CrowdRE. Although the ORKG project team has not yet consciously applied CrowdRE, they already address crucial parts of the CrowdRE cycle by motivating crowd members to participate, eliciting feedback, and monitoring context \& usage data, which they analyze to derive and implement the needs and requirements of the crowd.

To motivate crowd members, the project team uses established mechanisms and incentives to boost intrinsic and extrinsic motivation (see \ref{f1}). Feedback is elicited through channels integrated into the ORKG, i.a., Chatwoot, and standalone channels, i.a., a GitLab issue tracker (see \ref{f2}). However, analysis of this feedback is currently a weakness of the ORKG since the project team still does this manually (see \ref{f3}). While there are also basic mechanisms for monitoring context and usage data in place (see \ref{f4}), the feedback and the monitored data are currently not analyzed together. The overview and the findings show to what extent and how the ORKG already addresses CrowdRE and highlights gaps, which helps CrowdRE researchers make an informed decision on how suitable the ORKG is as a basis for future work and studies. Leveraging the inherent features of the ORKG for successful CrowdRE provides a solid basis for applying CrowdRE research in a real development setting, which in turn can help the project team further improve and enrich the platform and involve its crowd members in the development.

\begin{mdframed}
	\textbf{Answer to RQ1:}
	Although the ORKG project team has not yet consciously applied CrowdRE, the ORKG is a crowdsourcing platform that already has several important features for successful CrowdRE in place. Despite improvement potential\textemdash e.g., adding (semi-)automated feedback analyses\textemdash the ORKG provides a solid basis for researchers to apply and study CrowdRE in a real development setting and in close collaboration with the ORKG project team.
\end{mdframed}

In addition to reflecting on the ORKG's features, we also provided insights into our experiences using the ORKG as crowd members to communicate scholarly knowledge about CrowdRE research. Despite its limitations, especially regarding usability, the use of the ORKG and the exchanges with the project team were positive experiences. We observed that our feedback was received and addressed directly, highlighting the team's efforts to involve and acknowledge the crowd in the development. Overall, we achieved our goal of acquiring and curating the detailed results of two SLRs with a very different nature\textemdash Case~I focuses on qualitative and Case~II on quantitative scholarly knowledge. The ORKG has shown that it supports the acquisition and curation of both kinds of knowledge. Even though the input and representation of this information may initially appear to be human-readable only, the way the data is entered and stored in the underlying data structure makes the scholarly knowledge machine-actionable. As a result, scholarly knowledge, e.g., concepts and relationships, can be identified more easily due to greater terminological and conceptual precision and sharpness \cite{Auer.2018a}. In this way, the ORKG does not only provide researchers with another opportunity to publish and process literature, but also to develop novel services that make scholarly knowledge accessible with new search, retrieval, mining, and assistance applications \cite{Auer.2018a}. For example, the ORKG recently served as a data source for a dashboard that searches and visualizes academic literature on students' attitudes towards ICT in the PISA program\footnote[12]{\url{https://www.orkg.org/orkg/usecases/pisa-dashboard/}}. Our experiences have convinced us of the potential of the ORKG, and we propose its use as a platform for applying and communicating CrowdRE research to further advance this research while keeping its exponentially growing body of knowledge manageable \cite{Bornmann.2015}.

\begin{mdframed}
	\textbf{Answer to RQ2:}
	The ORKG enables a new way of communicating (CrowdRE) research through more comprehensive acquisition, curation, publication, and processing of scholarly knowledge than document-based works. For CrowdRE researchers, the ORKG does not only provide improved access to and communication of scholarly knowledge about their research, but also the opportunity to experience the ORKG as crowd members, offering a new perspective on their research.
\end{mdframed}

\section{Conclusion}
\label{sec:conclusion}
The ORKG is a crowdsourcing platform that already addresses crucial aspects of the CrowdRE cycle by offering several features that facilitate successful CrowdRE. However, these features must be further expanded to develop a systematic and holistic approach to achieve the goals of the ORKG project team, i.e., to involve researchers from various research fields as crowd members who both use the ORKG and participate in its development. This calls for collaborations between the ORKG project team and CrowdRE researchers to foster mutual benefits.
On the one hand, the ORKG project team will benefit from new crowd members and partners who can help to develop a corresponding systematic and holistic CrowdRE approach to continuously adapt the ORKG to the evolving needs and requirements of the crowd in the long term. On the other hand, CrowdRE researchers will benefit in two ways. First, they get a stable platform to apply their research that already has several features for successful CrowdRE, whose continuous development is guaranteed, and whose project team is interested in collaborating to implement CrowdRE in a real development setting. In this way, the ORKG can be an interesting research object in the future in terms of how CrowdRE techniques are incrementally added to the platform. Thus, the ORKG provides a basis for case studies in open source and open research settings (cf. Glinz \cite{Glinz.2019}).
Second, we laid the foundation for communicating CrowdRE research with the ORKG by acquiring, curating, and publishing scholarly knowledge about CrowdRE. CrowdRE researchers can build on this foundation and expand the ORKG by adding more CrowdRE papers with their contributions, as well as corresponding comparisons and smart reviews. In this way, they will also gain a fresh new perspective on their research as members of the ORKG's target crowd. For this reason, the ORKG is a crowdsourcing platform that can act as a useful multi-tool for CrowdRE research.

\section*{Acknowledgment}
This work was co-funded by the European Research Council for the project ScienceGRAPH (Grant agreement ID: 819536) and by the TIB -- Leibniz Information Centre for Science and Technology. We thank Sonnhild Namingha for proofreading this paper.

\bibliographystyle{IEEEtran}
\balance
\bibliography{IEEEabrv,references}

\end{document}